\documentclass[english]{article}
\usepackage[T1]{fontenc}
\usepackage[latin9]{inputenc}
\usepackage{amssymb}

\newcommand{\lyxaddress}[1]{
\par {\raggedright #1
\vspace{1.4em}
\noindent\par}
}

\usepackage{babel}

\begin{document}

\title{A NON \textendash{} SINGULAR COSMOLOGICAL MODEL WITH SHEAR AND ROTATION}

\author{G.K.Goswami$^{1}\&$ Mandwi Trivedi$^{2}$}

\maketitle

\lyxaddress{\begin{center}
$^{1}$Department of mathematics,
\par\end{center}}

\lyxaddress{\begin{center}
Kalyan PG College,Sector-7,
\par\end{center}}

\lyxaddress{\begin{center}
Bhilai Nagar 490006(C.G.)
\par\end{center}}

\lyxaddress{\begin{center}
$^{2}$Sri sankaracharya Engeering college,
\par\end{center}}

\lyxaddress{\begin{center}
Junvani,Bhilai(C.G.),
\par\end{center}}

\lyxaddress{\begin{center}
E-mail$^{1}$ gk\_goswami@yahoo.co.in.
\par\end{center}}
\begin{abstract}
We have investigated a non-static and rotating model of the universe
with an imperfect fluid distribution. It is found that the model is
free from singularity and represents an ever expanding universe with
shear and rotation vanishing for large value of time

Cosmology, non singular model 
\end{abstract}

\section{INTRODUCTION}

The \textquotedblleft{}Singularity theorems\textquotedblright{} of
Hawking, Penrose and Geroch$^{1}$show that any general relativistic
model universe (a) satisfying reasonable casuality and generality
condition (b) possessing a surface of the past (future) of which the
light cones starts converging and (c) containing matter which satisfies
the energy condition.

\begin{equation}
\left(T_{ij}-\frac{1}{2}Tg_{ij}\right)t^{i}t^{j}\geq0\end{equation}
 For any unit time like vector $t^{i}$ must contain a time like curve
with a past (future) end point at finite proper time (a singularity)
. In most known examples this end point is associated with an infinite
curuature singularity. The actual universe is expected to satisfy
generally all the above three requirements. Does this mean that the
universe actually passed through a singularity in the past ?

Some workers are resigned to the singularity and try to make it acceptable
by conjecturing that it is of the mild variety$^{2}$ or else by relegating
the big bang sin singularity to the \textquotedblleft{}infinite past\textquotedblright{}
according to physically appealing time scale$^{3}$ . Other contends
cosmological singularity is only a symptom of the incompleteness of
the theory. They propose that it is absent in more fundamental theory
in which Einstein\textquoteright{}s Equations are appropriately modified
by quantum effects$^{4}$ , phenomenological quardratic terms in the
curvature$^{5}$or the effect of the torsion$^{6}$. It is known that
once the energy condition (1) is given up, singularity free cosmologies
become possible. The Examples have been provided by Hoyle-Narlikar
C-field theory$^{7,12,13}$, Murphy$^{8}$, Fulling and parker$^{9}$,
Bekenstein$^{10}$and Prigogine$^{11}$. Recently Bianchi type II
and III vacuum cosmologies have been a matter of interest to the cosmologists$^{14-17}$.
In this paper we have investigated a non-singular and rotating model
of the universe by considering the following Bianchi type III general
metric 

\begin{equation}
ds^{2}=e^{2\psi}\biggl[-b(dx^{1})^{2}-e^{2x^{1}}a(dx^{2})^{2}-c(dx^{3})^{2}+2e^{x^{1}}dx^{4}dx^{2}+(dx^{4})^{2}\biggl]\end{equation}

Where $a,\, b\, c\,\&\,\psi$ are functions of time $x^{4}$, as a
generalization of Godel\textquoteright{}s stationary model$^{18}$
and subsequent rotating model given by Heckmann \textendash{} Schucking$^{19}$
and Reval \textendash{} Vaidya$^{20,21}$. The Energy momentum tensor
has been taken as that of imperfect fluid in Lichnerowicz$^{22}$.

\begin{equation}
T^{ij}=\left(p+\rho\right)u^{i}u^{j}-pg^{ij}+\left(q-p\right)v^{i}v^{j}\end{equation}

Where $\rho$ is the energy density of the fluid, p is the pressure
in x$^{1}$ or x$^{2}$ direction, q is the pressure in x$^{3}$ direction.
The fluid flow vectors u$^{i}$ is a normalized time \textendash{}
like vector, where as v$^{i}$ is a space \textendash{} like vector
so that

\[
u^{i}u_{i}=1\]
,

\[
u^{i}=\left(0,0,0,\frac{1}{\sqrt{g_{44}}}\right),a\]
,

\[
v^{i}v_{i}=-1\]

and

\[
v^{i}=\left(0,0,\frac{1}{\sqrt{-g_{33}}},0\right)\]

in co-moving co-ordinates.

By solving Einstein\textquoteright{}s field equations 

\begin{equation}
R^{ij}-\frac{1}{2}Rg^{ij}=-T^{ij}+\Lambda g^{ij}\end{equation}

It is found that the model is free from singularity and represents
an ever expanding universe with shear and rotation vanishing for large
value of time. The matter-density of the universe is positive through
out the span of the model but the pressure is found to be negative.
Thus like other non-singular cosmological models, this model also
violates energy-condition (1),In section (2), we have formed Einstein\textquoteright{}s
field equations (4), corresponding to metric (2) and Energy momentum
tensor (3). In section (3), a particular solution of Einstein\textquoteright{}s
field Equations have been obtained. We have also discussed various
geometrical and physical properties of the model. The last section
contains concluding remarks

\section*{2.EINSTEIN\textquoteright{}S FIELD EQUATION}

The non vanishing components of Ricci Tensor $R_{ij}$ for the metric
(2) are as follows.

$R_{11}=\left[-\left(\frac{ab_{4}}{2\left(a+1\right)}+\frac{ab\psi_{4}}{\left(a+1\right)}\right)_{4}-\left(\frac{ab_{4}}{2\left(a+1\right)}+\frac{ab\psi_{4}}{\left(a+1\right)}\right)\left(2\psi_{4}+\frac{a_{4}}{2\left(a+1\right)}-\frac{b_{4}}{2b}+\frac{c_{4}}{2c}\right)+\frac{2a+1}{2\left(a+1\right)}\right]$

$\vphantom{}$

$\vphantom{}$

$R_{22}=e^{2x^{1}}\left[-\left(\frac{a^{2}\psi_{4}}{\left(a+1\right)}+\frac{aa_{4}}{2\left(a+1\right)}\right)_{4}-\left(\frac{a^{2}\psi_{4}}{\left(a+1\right)}+\frac{aa_{4}}{2\left(a+1\right)}\right)\left(2\psi_{4}-\frac{\left(a+2\right)a_{4}}{2a\left(a+1\right)}+\frac{b_{4}}{2b}+\frac{c_{4}}{2c}\right)+\frac{a\left(2a+3\right)}{2b\left(a+1\right)}\right]$

$\vphantom{}$

$\vphantom{}$

$R_{33}=\left[-\left(\frac{ac_{4}}{2\left(a+1\right)}+\frac{ac\psi_{4}}{\left(a+1\right)}\right)_{4}-\left(\frac{ac_{4}}{2\left(a+1\right)}+\frac{ac\psi_{4}}{\left(a+1\right)}\right)\left(2\psi_{4}+\frac{a_{4}}{2\left(a+1\right)}+\frac{b_{4}}{2b}-\frac{c_{4}}{2c}\right)\right]$

$\vphantom{}$

$\vphantom{}$

$R_{42}=e^{x^{1}}\left[\left(\frac{a_{4}}{2\left(a+1\right)}+\frac{a\psi_{4}}{\left(a+1\right)}\right)_{4}+\left(\frac{a_{4}}{2\left(a+1\right)}+\frac{a\psi_{4}}{\left(a+1\right)}\right)\left(2\psi_{4}+\frac{a_{4}}{2\left(a+1\right)}+\frac{b_{4}}{2b}+\frac{c_{4}}{2c}\right)-\frac{1}{2b\left(a+1\right)}\right]$

$\vphantom{}$

\[
\vphantom{}\]

$R_{44}=$$\left(4\psi_{4}+\frac{a_{4}}{2\left(a+1\right)}+\frac{b_{4}}{2b}+\frac{c_{4}}{2c}\right)_{4}-\left(\frac{\left(a+2\right)\psi_{4}}{\left(a+1\right)}\right)_{4}$-$\left(4\psi_{4}+\frac{a_{4}}{2\left(a+1\right)}+\frac{b_{4}}{2b}+\frac{c_{4}}{2c}\right)\left(\frac{\left(a+2\right)\psi_{4}}{\left(a+1\right)}\right)$

$\vphantom{}$

$+\left(\psi_{4}+\frac{b_{4}}{2b}\right)^{2}+\left(\psi_{4}+\frac{c_{4}}{2c}\right)^{2}+\left(\frac{a}{a+1}\right)^{2}\left(\psi_{4}+\frac{a_{4}}{2a}\right)^{2}+\left(\frac{\left(a+2\right)\psi_{4}}{a+1}\right)^{2}+$$\frac{2\psi_{4}}{\left(a+1\right)}\left(\frac{a_{4}}{2\left(a+1\right)}+\frac{a\psi_{4}}{\left(a+1\right)}\right)-\frac{1}{2b\left(a+1\right)}$

$\vphantom{}$

$\vphantom{}$

$R_{41}=\left[\frac{-\psi_{4}}{a+1}+\frac{\left(2a+3\right)a_{4}}{4\left(a+1\right)^{2}}-\frac{\left(2a+1\right)b_{4}}{4\left(a+1\right)b}-\frac{c_{4}}{4\left(a+1\right)c}\right]$

$\vphantom{}$

$\vphantom{}$

$R_{21}=ae^{x^{1}}\left[\frac{\psi_{4}}{a+1}+\frac{\left(2+a\right)a_{4}}{4\left(a+1\right)^{2}a}-\frac{b_{4}}{4\left(a+1\right)b}-\frac{c_{4}}{4\left(a+1\right)c}\right]$

$\vphantom{}$

The Einstein\textquoteright{}s field equation (4), corresponding to
metric (2) and energy momentum tensor (3) yields

\begin{equation}
R_{21}=0\end{equation}

\begin{equation}
R_{41}=0\end{equation}

\begin{equation}
R_{44}=e^{-x^{1}}R_{42}\end{equation}

\begin{equation}
G_{11}=\left(p+\Lambda\right)g_{11}\end{equation}

\begin{equation}
G_{22}=-\left[\left(p+\rho\right)\frac{g_{42}^{2}}{g_{44}}-pg_{22}\right]+\Lambda g_{22}\end{equation}

\begin{equation}
G_{33}=\left(q+\Lambda\right)g_{33}\end{equation}

Equation (5) and (6) give

\begin{equation}
\frac{a_{4}}{a}=\frac{b_{4}}{b}\;(a+1\neq0)\end{equation}

and

\begin{equation}
4\psi_{4}=-\frac{a_{4}}{a\left(a+1\right)}-\frac{c_{4}}{c}\end{equation}

Equations (7) \textendash{} (10) give by using (11) and (12)

\begin{equation}
\frac{-a_{44}}{a+1}+\frac{a_{4}c_{4}}{2ac}-\frac{c_{4}^{2}}{4c^{2}}+\frac{\left(2a^{2}-6a-1\right)a_{4}^{2}}{4a^{2}\left(a+1\right)^{2}}=0\end{equation}

$\frac{a}{2\left(a+1\right)}\left[-4\psi_{44}-2\psi_{4}^{2}-\frac{2\left(a+2\right)a_{4}\psi_{4}}{a\left(a+1\right)}-\frac{2c_{4}\psi_{4}}{c}-\frac{a_{44}}{a}-\frac{c_{44}}{c}+\frac{c_{4}^{2}}{2c^{2}}+\frac{a_{4}^{2}}{2a\left(a+1\right)}-\frac{\left(a+2\right)a_{4}c_{4}}{2ac\left(a+1\right)}+\frac{1}{2ab}\right]=e^{2\psi}\left(p+\Lambda\right)$

\begin{equation}
\end{equation}

$\frac{-a}{2\left(a+1\right)}\left[-4\psi_{44}-2\psi_{4}^{2}-\frac{2\left(a+2\right)a_{4}\psi_{4}}{a\left(a+1\right)}-\frac{2c_{4}\psi_{4}}{c}-\frac{a_{44}}{a}-\frac{c_{44}}{c}+\frac{c_{4}^{2}}{2c^{2}}+\frac{a_{4}^{2}}{2a\left(a+1\right)}-\frac{\left(a+2\right)a_{4}c_{4}}{2ac\left(a+1\right)}+\frac{4a+1}{2ab}\right]=e^{2\psi}\left(\rho-\Lambda\right)$

\begin{equation}
\end{equation}

$\frac{a}{\left(2a+1\right)}\left[-4\psi_{44}-2\psi_{4}^{2}-\frac{2\left(a+2\right)a_{4}\psi_{4}}{a\left(a+1\right)}-\frac{2a_{4}\psi_{4}}{c}-\frac{2a_{44}}{a}+\frac{a_{4}^{2}}{2a^{2}}+\frac{a_{4}^{2}}{2a\left(a+1\right)}-\frac{\left(a+2\right)a_{4}^{2}}{2a^{2}\left(a+1\right)}-\frac{4a+3}{2ab}\right]=e^{2\psi}\left(q+\Lambda\right)$

\begin{equation}
\end{equation}

\section*{3. A PARTICULAR SOLUTION}

The functional form of $\psi$ may be determined if we make a simplifying
assumption that

\begin{equation}
\frac{a_{4}}{a}=\frac{c_{4}}{c}\end{equation}

This leads to this following solutions of equations (11) \textendash{}
(13)

\begin{equation}
b=\beta a\end{equation}

\begin{equation}
c=\gamma a\end{equation}

\begin{equation}
e^{4\psi}=\kappa^{4}\left(\frac{a+1}{a^{2}}\right)\end{equation}

and 

\begin{equation}
a_{4}=A\frac{\left(a+1\right)^{7/4}}{a}\end{equation}

where $\beta$ ,$\gamma$ , $\kappa$ and $A$ are arbitrary constants
of integration. We may replace

\begin{center}
$\beta x^{1}\rightarrow x^{1}\:\&\;\gamma x^{3}\rightarrow x^{3}$
\par\end{center}

Hence we take, without loss of generality

\[
a=b=c\]

Changing time t to proper time T via transformation 

\[
dT=e^{\psi}dt\]

and denoting new time derivative by a$_{T}$, the equation (19) is
transformed to

\begin{equation}
a_{T}=\frac{A\left(a+1\right)^{3/2}}{\kappa a^{1/2}}\end{equation}

which yields on integration

\begin{equation}
ln\left(a^{1/2}+\left(a+1\right)^{1/2}\right)-\frac{a^{1/2}}{\left(a+1\right)^{1/2}}=\frac{A}{2\kappa}T+L\end{equation}

where L is arbitrary constant of integration. The equation (23) shows
that for proper time T to be real

\[
a\geq0\]

And T increases monotonically with \textquotedblleft{}a\textquotedblright{}
.

Thus the metric (2) takes the following form

\begin{equation}
ds^{2}=dT^{2}+\frac{2\left(a+1\right)^{1/4}}{a^{1/2}}e^{x^{1}}dTdx^{2}-\left(a+1\right)^{1/2}\left[\left(dx^{1}\right)^{2}+e^{2x^{1}}\left(dx^{2}\right)^{2}+\left(dx^{3}\right)^{2}\right]\end{equation}

where we have taken arbitrary constant k = 1 for shake of simplicity.

\section*{4. Some Physical and geometrical properties of the model:--}

(i)The expansion scalar $\theta$is given by

\[
\theta=\frac{1}{3}u_{;k}^{k}=\frac{1}{12}A\frac{\left(3a-2\right)\left(a+1\right)^{1/2}}{a^{3/2}}\]

This shows that when a < 2/3, $\theta$ is negative, therefore the
model shows contraction for a <2/3 . However this can be avoided if
we choose the value of arbitrary constant L in equation (23) such
that when a = 2/3, proper time T = 0. 

This gives constant L = 0.111777. 

Thus for proper time T$\geq0$ , the model will represent an ever
expanding universe. For large value of \textquoteleft{}a\textquoteright{}
expansion scalar$\theta$ become stationary 

\[
a\rightarrow\infty\Rightarrow\theta\rightarrow\frac{1}{4}A\]

\begin{center}

\par\end{center}

(ii) Inserting value of a$_{4}$ and $\psi$ in the field equations
(14) \textendash{} (16) , we get \textendash{} 

\begin{equation}
p+\Lambda=-\frac{3}{16}A^{2}+\frac{1}{4\left(a+1\right)^{3/2}}\end{equation}

\begin{equation}
q+\Lambda=-\frac{3}{16}A^{2}+\frac{\left(4a+3\right)}{4\left(a+1\right)^{3/2}}\end{equation}

\begin{equation}
\rho-\Lambda=\frac{3}{16}A^{2}-\frac{\left(4a+1\right)}{4\left(a+1\right)^{3/2}}\end{equation}

\[
\frac{3}{16}A^{2}\geq\frac{4a+1}{4\left(a+1\right)^{3/2}}\Rightarrow\rho-\Lambda\geq0\]
.

As $\frac{4a+1}{4\left(a+1\right)^{3/2}}$ is decreasing function
of $a$ , its value is maximum at minimum value of a i.e. a=2/3.

This gives \[
\frac{3}{16}A^{2}\geq.426.\]

This choice of arbitrary constant A will make energy-density $\rho$
always positive throughout the span of the model, but then pressures
p and q will be negative barring the little span when

\[
\frac{3}{16}A^{2}\leq\frac{1}{4\left(a+1\right)^{3/2}}\;\;\&\;\;\frac{3}{16}A^{2}\leq\frac{4a+3}{4\left(a+1\right)^{3/2}}\]

The negative value is very little nearly -.426. 

(iii) Equation (1), in presence of $\Lambda$ should be read as

\[
(T_{ij}-\frac{1}{2}Tg_{ij}+\Lambda g_{ij})t^{i}t^{j}\geq0\]

If we take $t^{i}=u^{i}$ as a special case, then the equation turns
into 

\[
\rho+2p+q+2\Lambda\geq0\]

in our case. Inserting values of $\rho$, p and q from equations (25-27)
we get

\[
\rho+2p+q+2\Lambda=\frac{-3}{16}A^{2}+\frac{1}{\left(a+1\right)^{3/2}}\]

which is negative, although it is nearly -.213. As stated earlier
this is the requirement of non gingular models which violates conditions
laid by Hawking, Penrose and Geroch$.^{1}$

(iv) The non vanishing components of shear tensor 

\[
\sigma_{ij}=u_{i;j}-\left(g_{ij}-u_{i}u_{j}\right)\theta\]

\[
\sigma_{11}=-A\frac{\left(a+1\right)}{a^{3/2}}\]

\[
\sigma_{22}=Ae^{2x^{1}}\frac{\left(a-2\right)\left(a+1\right)}{12a^{1/2}}\]

\[
\sigma_{33}=-A\frac{\left(a+1\right)}{6a^{3/2}}\]

$\because$shear\textendash{} scalar

\[
\sigma^{2}=\frac{1}{2}\sigma_{ij}\sigma^{ij}=A^{2}\frac{\left(3a^{2}+4a+4\right)}{96\left(a+1\right)a^{3}}\]

\[
\because a\rightarrow\infty\Rightarrow\sigma^{2}\rightarrow0\]

i.e. for large values of time, shear vanishes.

(v) The non-vanishing component of angular velocity vector 

$ $\[
\omega^{i}=\frac{1}{2}\sqrt{\left(-g\right)}\epsilon^{ijkl}\left(u_{j;k}-u_{k;j}\right)u_{l}\]

is given by

\[
\omega^{3}=\frac{1}{2\left(a+1\right)}\]
 \[
\because a\rightarrow\infty\Rightarrow\omega\rightarrow0\]

This shows that the model possesses a non zero finite rotation about
x$^{3}$- direction, but it goes on diminishing monotonically with
the passage of time.

\section*{5. Conclusion}

The main purpose of the paper is to investigate Cosmological model
without singularity within frame work of general relativity. We have
proposed an ever expanding and nonsingular model of rotating universe.
As stated in the introduction that the singularity which appears in
almost all the standard Cosmological models of the universe is not
accepted by workers in the field (see ref.{[}2-13{]}) this paper is
a simple and elegant effort in this direction. The initial state of
the universe was  inhomogeneous and anisotropic, but slowly-slowly
with the advent of time inhomogenity and anisotropy ceased out and
at present the universe is spatially homogeneous and isotropy. This
behavior is very well depicted in our model. We started with universe
filled with imperfect fluid showing shear and rotation but these elements
are gradually decreasing with advent of time and ultimately the model
will be turned up into a spatially homogeneous and isotropic one with
constant values of density, pressure and expansion scalar

It is noteworthy to state that our model is not a special case of
rotating model given in the reference {[}19-21{]}. These models are
valid only for a limited time interval whereas our model is ever lasting.

\section*{6. Acknowledgement}

The author is highly grateful to Prof. V.B. Johri, emeritus Prof.,
Deptt. of Mathematics and Astronomy Lucknow University for his valuable
suggestions and comments during the preparation of the paper.


\begin{thebibliography}{22}
\bibitem{key-3} R. Penrose and S.W. Hawking, Proc. R. Soc. A314,
529 (1970) ; R.P. Geroch, Phys. Rev. Lett. 17, 445 (1966) .

\bibitem{key-4} G.F.R. Ellis and A.R. King, Commun. Math. Phys. 38
, 119 (1974).

\bibitem{key-5} C.W. Misner, Phys. Rev. 186, 1328 (1969).

\bibitem{key-7} J.A. Wheeler in Relativity, Groups and Topology,
edited by B.S. De witt (Gordon and Breach, New York, 1964) ; V.L.
Ginzburg, Comments Astrophys. Space Phys. 3 , 7 (1971).

\bibitem{key-9}H. Nariai and K. Tomita, Prog. Theor. Phys. 46 , 776
(1971).

\bibitem{key-10} A. Trautman, Nat. Phys. Sci. 242 , 7 (1973) ; J.
Taffel , Phys. Lett. 45A , 341 (1973). However, see J. Stewart and
P. Hajicek, Nat. Phys. Sci. 244 , 96 (1973).

\bibitem{key-11} F. Hoyle and J.V. Narlikar (1963) , Proc. Roy. Soc.
A. , 273 , 1.

\bibitem{key-15} G.L. Murphy , Phys. Rev. D 8, 4231 (1973).

\bibitem{key-12} S. Fulling and L. Parker, Phys. Rev. D 7 , 2357
(1973).

\bibitem{key-17} Jecob D. Bekenstein Phys. Rev. D11 , 2073 (1975).

\bibitem{key-18} Progogine GRG vol.21, No 8, 1989.

\bibitem{key-19} F Hoyle , G . Burbidge and J V Narlikar \textquotedblleft{}
A quasi- steady state cosmological model with creation of matter A
p Journal p 410.43

\bibitem{key-20}F Hoyle , G . Burbidge and J V Narlikar 2000 Adifferent
approach to cosmology(Cambridge Univ. Press.) 

\bibitem{key-21} T. Christodoulakis \& others J.Math.Phys,42,8,3580-3608(2001) 

\bibitem{key-22} T. Christodoulakis \& others Commun. Math. Phys.
226,377(2002) 

\bibitem{key-23} T. Christodoulakis \& others J Phys. A 36,427(2003)

\bibitem{key-25}T. Christodoulakis \& others J.Math.Phys,47,102502(2006)

\bibitem{key-26}K. Godel (1949) Rev. Mod. Phys. 21 , 447. . 

\bibitem{key-27} O. Heckmann and E. Schucking (1958) , Reports Solvay
Conference, Bruxelles.

\bibitem{key-28} H.M. Raval and P.C. Vaidya- current science, 34
, 17 (1965)

\bibitem{key-30}H.M Raval and P.C. Vaidya- Ann. Inst. Henri Poincare
4 , 21(1966) 

\bibitem{key-31} A.Lichnerouriz(1955) Theories Relativistics de la
Gravitation et l\textquoteright{}electromagentisme\textquoteright{}
, (mason Pub.) , p 14.
\end{thebibliography}
\end{document}